\documentclass[aps,pra,a4paper,twocolumn]{revtex4-1}
\usepackage[utf8]{inputenc}
\usepackage[T1]{fontenc}
\usepackage{amsmath,amsthm,amssymb}
\usepackage{bbm}

\usepackage{graphicx}
\usepackage{tikz}
\usepackage{braket}
\usetikzlibrary{decorations.pathmorphing}
\usetikzlibrary{arrows}
\usetikzlibrary{matrix}

\usepackage{color}
\usepackage[normalem]{ulem}
\definecolor{gray}{rgb}{.6,.6,.6}
\definecolor{darkyellow}{rgb}{.6,.5,0}
\definecolor{darkgreen}{rgb}{0,.6,0}
\definecolor{darkblue}{rgb}{0,0,.6}

\usepackage{hyperref}
\usepackage{verbatim}

\newcommand{\tr}[2][]{\text{Tr}_{ #1 } ( #2 )}

\begin{document}

\title{Conditioned spin and charge dynamics of a single electron quantum dot}

\author{Eliska Greplova $^{1}$, Edward A. Laird $^{2}$, G. Andrew D. Briggs $^{2}$, Klaus Mølmer $^{1}$ }
\affiliation{$^{1}$ Department of Physics and Astronomy, Aarhus University, Aarhus, Denmark}
\affiliation{$^{2}$ Department of Materials, Oxford University, Oxford, UK}

\date{\today}

\begin{abstract}
In this article we describe the incoherent and coherent spin and charge dynamics of a single electron quantum dot. We use a stochastic master equation to model the state of the system, as inferred by an observer with access to only the measurement signal. Measurements obtained during an interval of time contribute, by a past quantum state analysis, to our knowledge about the system at any time $t$ within that interval. Such analysis permits precise estimation of physical parameters, and we propose and test a modification of the classical Baum-Welch parameter re-estimation method to systems driven by both coherent and incoherent processes.
\end{abstract}

\maketitle

\section{Introduction}

\begin{figure}
\includegraphics[scale=0.53]{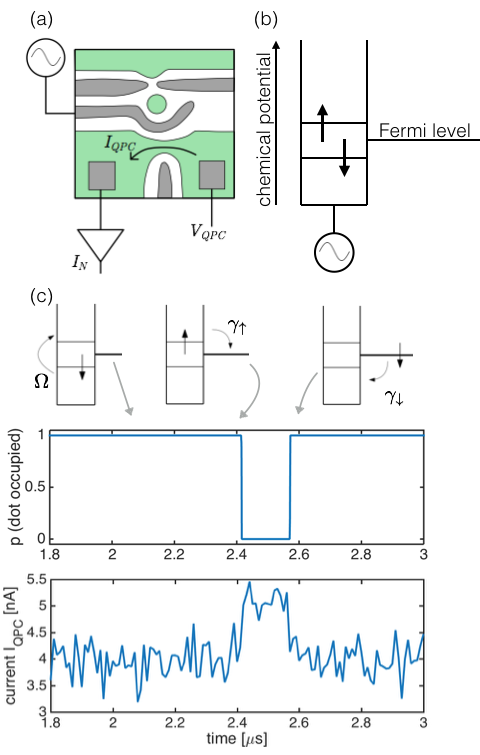}
\caption{(a) Schematic of the experimental setup: GaAs quantum dot with the quantum point contact, yielding the experimentally accessible current $I_{QPC}$; (b) Energy levels being populated through the exchange of electrons with the the lead electrode: a spin-down (up) electron can tunnel into (off) the dot; (c) Relation of the system dynamics to the measured current: the Rabi precession does not lead directly to a change in current, but does enable the electron to leave the dot, causing a temporary current increase until another electron tunnels onto the quantum dot.}
\label{QDFig}
\end{figure}

In quantum mechanics, the state of a physical system is described by a  wave function $\psi$ or a density  matrix $\rho$ which provides the probability of the outcomes of any measurement that we might carry out on the system. Following such a measurement, the state of the system changes as described by the formalism of projective measurements \cite{von1955mathematical} and its extension to more general measurements \cite{nielsen2010quantum} and to continuous monitoring of quantum systems \cite{WiMi2010, JacobsSteck}. 

The information about the system from stochastic measurement signals can be used for the purpose of state reconstruction \cite{PhysRevLett.72.3439, PhysRevLett.98.090401, PhysRevLett.97.180403, PhysRevLett.105.150401} and precision measurements of physical parameters \cite{PhysRevA.69.032109, PhysRevA.64.042105, PhysRevA.87.032115,PhysRevA.89.043839, PhysRevA.60.2700}. In this work we achieve this by a a combination of Bayesian analysis and modified Baum-Welch re-estimation that extracts the quantum state and the physical parameters governing the system dynamics from the measurement data. We apply this to the single-electron occupation of a quantum dot. Because the occupation depends on quantum tunnelling which in turn depends on the spin state \cite{RevModPhys.79.1217}, we have a combination of incoherent and coherent dynamics. This requires a modification to the conventional Baum-Welch estimation scheme, with repeated application until all the parameters have converged.

We consider a new scheme for repeated and continuous monitoring of a spin qubit that tunnels on and off a quantum dot~\cite{RevModPhys.79.1217}. A static magnetic field splits the spin-up and spin-down state energies, such that only a spin-down electron may tunnel into the dot and prevent further charging of the dot by Coulomb blockade. A resonant drive causes oscillations between the spin-up and spin-down states. There is therefore a mixture of coherent spin dynamics and incoherent tunneling events, as the Rabi oscillations are interrupted when the excited spin-up electron tunnels out of the dot. A quantum point contact (QPC) which transmits an electron current that depends on the charge but is insensitive to the spin dynamics on the quantum dot is used to continuously monitor the electron tunneling  dynamics~\cite{Elzerman2004,Petta2180,QPCranieri,QPCcahay}. Our theory will apply to the analysis of real experimental data, but in this work it will be illustrated on a simulated system dynamics, where the tunneling events in and out of the quantum dot occur governed by a stochastic master equation \cite{WiMi2010}. We note that even though our experiments are not sensitive to the electron spin state, the Rabi oscillatory dynamics will reveal itself through the distribution of time intervals spent on the dot, since the electron enters and leaves in different spin states. We thus aim to recover the quantum state and the physical parameters governing the electron spin and charge from the stochastic measurement signal of the charge sensor. We show how the time dependent dot occupation can be estimated and how this estimate is improved by incorporating information from the subsequent sensor record. Finally, we show how the spin precession becomes imprinted on the charge dynamics, and we present a method to estimate efficiently the parameters of the qubit Hamiltonian from the noisy charge measurement record.

The article is organized as follows: In Sec. II, we introduce the master equation and quantum trajectory description of the system subject to ideal probing of the charge state. In Sec. III, we present the case of continuous probing of the charge state, and illustrate the discrepancy between the true and the estimated state of the system. In Sec. IV, we show how the past quantum state formalism employs the full time dependent signal and provides an estimate of the time dependent charge state in better agreement with the true evolution of the system. In Sec. V, we propose a combination of Bayesian analysis and modified Baum-Welch re-estimation that extracts the physical parameters governing the system dynamics from the measurement data. In Sec. VI, we present a conclusion and outlook.

\section{Coherent and incoherent processes, conventional and stochastic master equation}
\begin{figure}
 \centering
\includegraphics[scale=0.28]{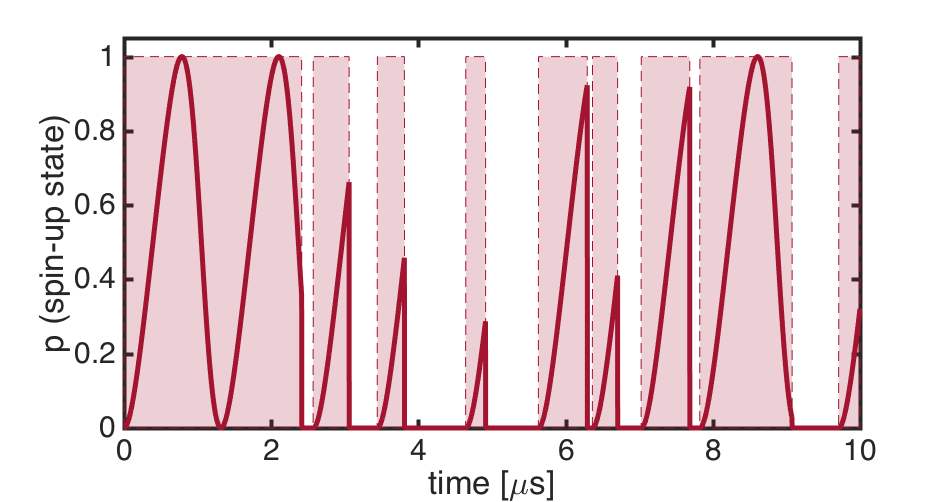}
 \caption{Example of a particular quantum trajectory: the occupation of the spin-up state is plotted as a function of time and shows the Rabi precession of the spin interrupted by the quantum jumps where the electron leaves the dot. The shaded red regions correspond to the time intervals when the quantum dot is occupied irrespective of the spin state. The figure is obtained with a spin Rabi frequency of $\Omega=5$ MHz, and tunneling rates $\gamma_\uparrow=\gamma_\downarrow=3$ MHz.}
 \label{fig:trajectory}
 \end{figure}

We consider a gate-defined quantum dot that can contain zero or one electron (Fig. 1(a)). The quantum dot is coupled to an electron reservoir such that electrons can tunnel on and off. The three states available are the state $\ket{0}$ with no electron charge on the quantum dot, and the singly charged spin-up and spin-down states $\ket{\uparrow}$ and $\ket{\downarrow}$. The dot is measured via a quantum point contact (QPC) acting as a charge sensor. The current $I_{QPC}$ through this point contact is sensitive to the charge occupation; when the dot is empty, the average current is $I_0$, while, when the dot is occupied with one electron, the current is $I_1$.

We suppose that the device is placed in a magnetic field to introduce a Zeeman splitting of the two spin states, adjusted using gate voltages so that the spin-up and spin-down levels straddle the Fermi level of the reservoir (Fig. \ref{QDFig}b). Electron tunneling is now spin-sensitive; if the quantum dot is charged with an electron in state $\ket{\uparrow}$, it will tunnel off the dot with rate $\gamma_{\uparrow}$; if the quantum dot is empty, a spin-down electron will tunnel onto the dot with rate $\gamma_{\downarrow}$. A coherent drive at the qubit resonance (Larmor) frequency induces coherent precession at the Rabi frequency $\Omega$, and when the device evolves between the three states, the charge state sensitive $I_{QPC}$  fluctuates around different characteristic values as shown in Fig. \ref{QDFig}c.

The Hamiltonian describing the resonant drive between the up and down states reads $(\hbar=1)$
\begin{equation}
\label{eq:hamil}
H=\frac{\Omega}{2}(\ket{\uparrow}\bra{\downarrow}+\ket{\downarrow}\bra{\uparrow}),
\end{equation}
with the Rabi frequency $\Omega$.

Let us define the operators associated with the incoherent transfer of the electron between the dot and the electron states in the Fermi sea:
\begin{align}
&c_{\downarrow}=\ket{0}\bra{\downarrow},\quad c^{\dagger}_{\downarrow}=\ket{\downarrow}\bra{0},\\
&c_{\uparrow}=\ket{0}\bra{\uparrow},\quad c^{\dagger}_{\uparrow}=\ket{\uparrow}\bra{0}.
\end{align}
If $\gamma_{\downarrow},\gamma_{\uparrow}$ denote the respective tunneling rates and $dt$ is an infinitesimal time interval, the master equation \cite{WiMi2010} for the system reads
\begin{align}
\label{eq:MEq1}
\frac{d\rho}{dt}&= -i[H,\rho]+\frac{\gamma_{\downarrow}}{2}(2 c_{\downarrow}^{\dagger}\rho c_{\downarrow}-c_{\downarrow}c_{\downarrow}^{\dagger}\rho-\rho c_{\downarrow}c_{\downarrow}^{\dagger}) \\ \nonumber
&+\frac{\gamma_{\uparrow}}{2}(2c_{\uparrow}\rho c_{\uparrow}^{\dagger} - c_{\uparrow}^{\dagger}c_{\uparrow}\rho-\rho c_{\uparrow}^{\dagger}c_{\uparrow}).
\end{align}
This master equation describes the average dynamics of the unobserved system, subjected to both coherent driving between the spin states and incoherent tunneling onto and off the dot, described by rate equation terms. After a time of a few $\gamma_{\uparrow,\downarrow}^{-1}$ this equation causes $\rho$ to converge to a steady state density matrix with populations of the three states and coherences between the two occupied spin states. Here, the system is modelled at zero temperature, but finite temperature can be straightforwardly added into the master equation \cite{gardiner2004quantum}. The typical temperature in the dilution fridge is below $30$ mK, while the energy of Zeeman splitting of the electron spin resonance at typical magnetic fields corresponds to a temperature of $\approx 300$ mK, which means that for our purposes the thermal bath of the quantum dot is effectively at zero temperature.

If we imagine that we could monitor the occurrence of each tunneling event, Eq.\eqref{eq:MEq1} would be replaced by a stochastic quantum trajectory. This means that in each time step, $dt$, the system density matrix is first evolved as
\begin{align}
\label{eq:MEq2}
\frac{d\rho}{dt}&= -i[H,\rho]-\frac{\gamma_{\downarrow}}{2}(c_{\downarrow}c_{\downarrow}^{\dagger}\rho+\rho c_{\downarrow}c_{\downarrow}^{\dagger}) \\ \nonumber
&-\frac{\gamma_{\uparrow}}{2}(c_{\uparrow}^{\dagger}c_{\uparrow}\rho+\rho c_{\uparrow}^{\dagger}c_{\uparrow})
\end{align}
and then made subject to one of the quantum jumps onto or off the quantum dot:
\begin{equation*}
\rho \rightarrow \frac{c_{\downarrow}^{\dagger}\rho c_{\downarrow}}{\tr{c_{\downarrow}^{\dagger}\rho c_{\downarrow}}} = |\downarrow\rangle\langle \downarrow |
\end{equation*}
or,
\begin{equation*}
\rho\rightarrow  \frac{ c_{\uparrow}\rho c_{\uparrow}^{\dagger}}{\tr{ c_{\uparrow}\rho c_{\uparrow}^{\dagger}}}
= |0\rangle\langle 0|.
\end{equation*}
These jumps occur with the probabilities $dp=\tr{c_{\downarrow}^{\dagger}\rho c_{\downarrow}} \gamma_{\downarrow}dt$ and $dp=\tr{c_{\uparrow}\rho c_{\uparrow}^{\dagger}} \gamma_{\uparrow}dt$, respectively, and $\rho$ is  renormalized, before the evolution is continued in the following time step.

 This stochastic evolution forms a quantum trajectory \cite{carmichael1993open}, corresponding to the quantum dynamics conditioned on the hypothetical, perfect probing of the tunneling dynamics. We shall thus use such a trajectory dynamics to simulate the system, see Fig.\ref{fig:trajectory} which shows a sample evolution obtained in the way just described. The curve in the figure shows the spin-up population, dropping discontinuously to zero in connection with the simulated tunneling events off the quantum dot. The total occupation of the dot is unity in the intervals with the temporally modulated spin up population.

\section{Continuous probing of the charge dynamics}

\begin{figure}
 \centering
 \includegraphics[scale=0.27]{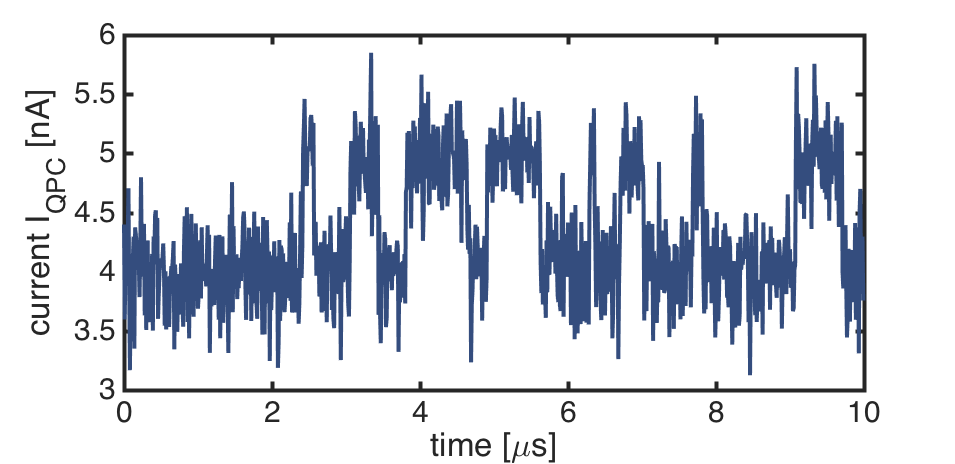}
 \caption{Simulation of the measured current through the QPC with the same parameters as in Fig.2, and with QPC current rates $r_0=31.21$ GHz and $r_1=24.97$ GHz. These rates lead to the charge dependent currents $I_{QPC,0}=5$ nA and $I_{QPC,1}=4$ nA while Poissonian count statistics (see text) leads to the standard deviation $\sigma_0=0.28$ nA and  $\sigma_1=0.26$ nA for a measurement binning time of $10$ ns.}
 \label{fig:current}
 \end{figure}

\begin{figure}
\centering
\includegraphics[scale=0.5]{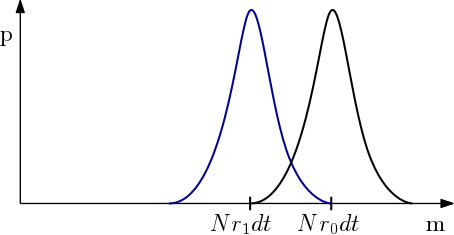}
\caption{The probability that $m$ electron counting events are registered by the QPC during a measurement time $T=Ndt$ (where N is large). For the empty and the charged quantum dot, a perfect detection yields QPC counts that are Poisson distributed with the mean values $Nr_0dt$ and $Nr_1dt$ and standard deviations $\sigma_0=\sqrt{Nr_0dt}$ and $\sigma_1=\sqrt{Nr_1dt}$ respectively.}
\label{fig:poisson}
\end{figure}

We now describe the information about the quantum state available to an experimentalist who has access only to the fluctuating QPC current signal  as illustrated in Fig. \ref{fig:current}. We shall then compare this information with the simulated record of occupations on and off the quantum dot, depicted in Fig. \ref{QDFig}, that gives rise to the QPC signal variation.

For the purpose of this analysis, we consider the current through the charge sensor as a stochastic counting signal, with two possible electron counting rates $r_{0,1} \equiv I_{0,1}/e$ depending on the charge on the dot.
We assume a $1$ nA difference between $I_0$ and $I_1$. While the QPC current record is measured as a continuous signal and does not resolve individual electrons, it is convenient to describe the measurement back action by the positive operator-valued measure (POVM) formalism \cite{nielsen2010quantum}, i.e., by operators $M_c$ and $M_{nc}$ yielding the probabilities and the back action associated with the count and no-count of each single electron passing the QPC in a time interval $dt$:
\begin{align}
\begin{split}
 \label{povms}
M_c &= \sqrt{r_0 dt} \Pi_0 + \sqrt{r_1 dt}\Pi_1,\\
M_{nc} &= \sqrt{1-r_0 dt} \Pi_0 + \sqrt{1- r_1 dt}\Pi_1,
\end{split}
\end{align}
where $\Pi_0 = |0\rangle\langle 0|$ is the projection on the empty dot state and $\Pi_1=|\uparrow\rangle\langle\uparrow| + |\downarrow\rangle\langle\downarrow|$ is the projection operator on the charged dot state, with eigenvalues $0$ and $1$.

The probability for an electron to tunnel through the QPC in interval $dt$ (a 'click event') is
\begin{equation} \label{povm-prob}
P_{click} = \tr{M_c \rho M_c^\dagger} = \rho_{00} r_0 dt + (\rho_{\uparrow \uparrow} + \rho_{\downarrow \downarrow}) r_1 dt
\end{equation}
and the state conditioned on the click or no click ($x=c$ or $x=nc$), is
\begin{equation} \label{povm-back}
\rho|_x = (M_x \rho M_x^{\dagger})/\tr{M_x \rho M_x^{\dagger}}.
\end{equation}

In a real experiment one measures a macroscopic fluctuating current corresponding to the large number of electrons, $m$, passing through the QPC in finite time increments $\Delta t=Ndt$.
If the system is subject to no further evolution, Eqs. (\ref{povms}-\ref{povm-back}) then lead to an integrated number of electrons going through the QPC given by a sum of two Poisson distributions with mean values $N r_0 dt$ and $N r_1 dt$ and standard deviations $\sigma_0=\sqrt{Nr_0dt}$ and $\sigma_1=\sqrt{Nr_1dt}$ respectively (see Fig. \ref{fig:poisson}). In the limit of large $N$, the current fluctuations can be modelled with Gaussian noise and various corrections to Poissonian counting statistics \cite{PhysRevB.90.205429} may be incorporated by adjusting the Gaussian widths. To simplify the presentation, we restrict our analysis to the Poissonian case given by Eqs. (\ref{povms}-\ref{povm-back}).
As the descriptions yield equivalent result, we shall refer to the conceptually simpler POVM formalism throughout this work, while stochastic differential equations may be more efficient in some numerical applications.

\begin{figure}
 \centering
\includegraphics[scale=0.33]{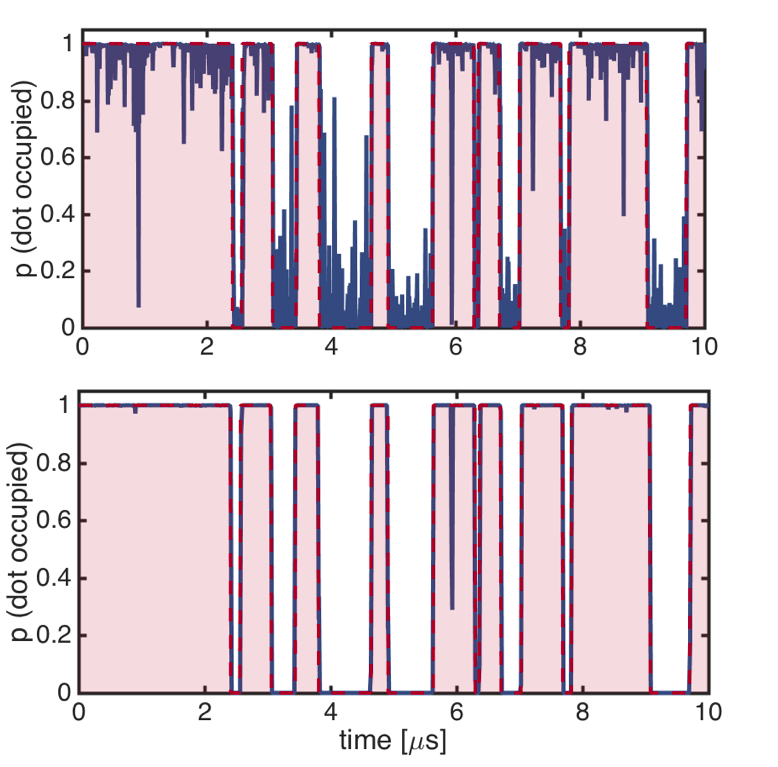}
 \caption{Comparison of true and inferred dynamics: the blue solid curve shows the occupation of the quantum dot as inferred from the QPC current. The red regions show the time intervals where the dot is actually occupied according to the simulated tunneling events on and off the quantum dot. The upper panel shows this comparison when the probability from the QPC current is inferred using the conditioned master equation. In the lower panel the probability is calculated using the past quantum state method.}\label{fig:occupation}
 \end{figure}

We shall use the symbol $\rho$ to represent the density matrix inferred from the QPC measurements, which differs from the "true" simulated state $\rho_{true}$. The QPC counting rates and the noisy signals shown in Fig. \ref{fig:current} are determined from $\rho_{true}$. Then we use only the QPC measurement current, and knowledge about the spin rotation Hamiltonian and the tunneling rates on and off the dot. We apply Eq.\eqref{eq:MEq1} and the POVM operators associated with the QPC detection to evolve the density matrix $\rho$. Note that, in general, the simulated $\rho_{true}$ should also incorporate the back action of the QPC probing in its evolution, but as $\rho_{true}$ populates with certainty either the charged state or the uncharged state, it is invariant under the QPC back-action Eq. \eqref{povm-back}. In the upper panel of Fig. \ref{fig:occupation} we compare the quantum dot occupation according to $\rho$ (blue curve) and $\rho_{true}$ (red regions). With the chosen parameters, the QPC probing is able to follow the charge dynamics of the quantum dot, but the state assignment is subject to statistical fluctuations due to the measurement noise.

\section{Past Quantum State}

For the parameters used here, the charge state changes on a similar time scale as our acquisition of measurement data, and as Fig. \ref{fig:occupation} shows, we cannot track the true system evolution perfectly because of the noise of the QPC current. In particular, we observe sharp features, indicating tunneling events, while subsequent rapid return to the original state, reveals that it was most probably a random signal fluctuation rather than a change of state. This is a well known problem in classical inference and has led to the introduction of data smoothing algorithms which incorporate the full signal and use data obtained both before and after $t$ in the analysis of the state of a system at the time $t$. This can be done in a rigorous analysis known as the "forward-backward" analysis for classical hidden Markov models \cite{PhysRevA.89.043839} and as the past quantum state (PQS) \cite{PhysRevLett.111.160401,xu2015correlation, PhysRevA.94.042334} and the quantum smoothing \cite{PhysRevLett.115.180407} formalism for quantum systems.

As the quantum formalism deals with the assignment of probabilities to measurement outcomes, we can present our knowledge about a quantum system at time $t$ by a general expression for such probabilities. The most general  measurements are described by the formalism of POVMs \cite{nielsen2010quantum}. We already saw examples of this formalism in our treatment of the QPC counting signal, and it quite generally assigns to any measurement a set of operators $\{M_i\}$, with $\sum_i M_i^\dagger M_i = I$, each representing an outcome ($i$) of the measurement.

The probability for outcome $i$ is conventionally given by the density matrix expression,
\begin{align}
P(i)=\frac{\tr{M_i\rho M_i^\dagger}}{\sum_i\tr{M_i\rho M_i^\dagger}},
\end{align}
where the density matrix $\rho$ may be the solution to our stochastic quantum dynamics described above, i.e., $\rho(t)$ is conditioned on the QPC  measurement outcomes that occurred and were read out until time $t$.

The past quantum state formalism \cite{PhysRevLett.111.160401} offers an improved expression that incorporates the subsequent signal record via an ``effect matrix'' $E(t)$. In this formalism, the probability that a measurements of the observable corresponding to the POVM  $\{M_i\}$ yielded the outcome $i$ is
\begin{equation}
\label{eq:genBorn}
P_{PQS}(i)=\frac{\tr{M_i\rho(t) M_i^{\dagger}E(t)}}{\sum_{i}\tr{ M_i\rho(t) M_i^{\dagger}E(t)}}.
\end{equation}
Here, $\rho(t)$ depends only on the QPC measurement outcomes obtained \textit{before} $t$ and the matrix $E(t)$ depends only on the QPC measurement outcomes obtained
\textit{after }time $t$. In particular, $E$ follows from an adjoint master equation that is solved backwards in time. The expression Eq.~(\ref{eq:genBorn}) follows from the quantum theory of measurements and conditional probabilities, and it has been applied to experiments with Rydberg atoms in microwave cavities \cite{PhysRevA.91.062116} and with superconducting qubits \cite{PhysRevLett.114.090403,PhysRevLett.112.180402,PhysRevA.94.050102}, where its predictions have been confirmed and used to identify quantitative properties of the system.

To evaluate the matrix $E$ introduced in \cite{PhysRevLett.111.160401}, we assume the final value $E=I$ in the future of all probing measurements, and propagate the matrix elements backwards in time by incorporating the known Hamiltonian evolution and master equation damping terms, a well as the QPC measurement outcomes, in a way that is  adjoint to the evolution of $\rho$ \cite{PhysRevLett.111.160401}. Between QPC counting events, the matrix $E$ pertaining to the experimental observer, obeys the equation
\begin{align}
&\frac{E(t-dt)-E(t)}{dt}=i[H,E]+\frac{\gamma_{\downarrow}}{2}(2 c_{\downarrow}E c_{\downarrow}^{\dagger}-c_{\downarrow}c_{\downarrow}^{\dagger}E-E c_{\downarrow}c_{\downarrow}^{\dagger})\label{eq:SMEE}  \nonumber \\
&+\frac{\gamma_{\uparrow}}{2}(2c_{\uparrow}^{\dagger}E c_{\uparrow} - c_{\uparrow}^{\dagger}c_{\uparrow}E-E c_{\uparrow}^{\dagger}c_{\uparrow}).
\end{align}
In addition to the dynamics, decribed by Eq. \eqref{eq:SMEE}, $E$ undergoes changes associated with the QPC probing of the system. This is described by and is analogous to the POVM operations acting on $\rho$ in Eqs.~ \eqref{povms}-\eqref{povm-back}, $E|_x = (M_x^{\dagger} E M_x)/\tr{M_x^{\dagger} E M_x}$).
We note that unlike the master equation Eq. \eqref{eq:MEq1} the evolution Eq. \eqref{eq:SMEE} does not preserve the trace of $E$, but the explicit normalization of the outcome probabilities ensures
the physical applicability of Eq. \eqref{eq:genBorn}.

The results of the PQS analysis are depicted in the lower panel Fig. \ref{fig:occupation}, again, in comparison with the simulated `true' state. We observe that the assignment of charge states to the dot is significantly improved compared to the upper panel in Fig. \ref{fig:occupation}.

\section{Parameter estimation}
\label{sec:estimation}

\begin{figure}
 \centering
 \includegraphics[scale=0.35]{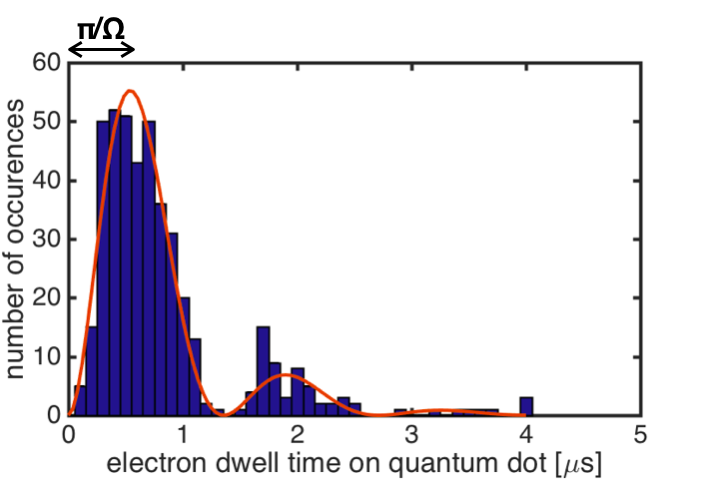}
 \caption{Histogram of the time intervals in which the quantum dot is occupied by an electron according to the past quantum state analysis of the QPC measurement data. The modulation reflects the Rabi oscillation between the spin states and the red curve represents the fit of the histogram by the expression \eqref{eq:OmegaFit} which leads to an estimated Rabi frequency of $\sim5.2$~MHz in reasonable agreement with the value of 5MHz used in our simulation.} \label{fig:histogram}
 \end{figure}

In order to infer the time dependent occupation of the dot from the measurement signal, we solved the master equation for $\rho(t)$ and the adjoint equation for $E(t)$, taking into account the measurement data, which could come from a real experiment but was synthesized by simulating the dynamics of a "true state". While the master equation \eqref{eq:MEq1} contains terms governed by the numerical values of the Rabi frequency $\Omega$ and the tunneling rates $\gamma_\downarrow$, $\gamma_\uparrow$, the abrupt changes of the occupation are mainly governed by the random measurement back-action, cf. the variations in the current signal shown in Fig.~\ref{fig:current}.

Larger or smaller values of the Rabi frequency and the rates would not be in conflict with short segments of the signal and the inferred occupation dynamics, but we expect a statistical agreement with the physical parameters when the system is observed over longer times. Thus, the periods with no occupation of the dot should follow an exponential distribution with the rate constant $\gamma_{\downarrow}$, and since the electron always enters the quantum dot in the spin down state and leaves in the spin up state, we expect to see the intervals of occupation of the dot cluster around odd multiples of  $\sim \pi/\Omega$, and there should be very few intervals close to even multiples of $\sim \pi/\Omega$.

This intuition is confirmed by making a histogram of the electron dwell times on the dot, extracted from the PQS analysis of the dot occupation (Fig.~\ref{fig:histogram}). 
Even more so, it permits a fit to the analytical dwell time distribution for different values of $\Omega$,
\begin{equation}
\label{eq:OmegaFit}
w(t)=-\frac{2\Omega^2\gamma_\uparrow}{\kappa^2}\exp\left(-\frac{t\gamma_\downarrow}{2}\right)\left(\cos\left(\frac{t\kappa}{2}\right)-1\right),
\end{equation}
where $\kappa=\sqrt{4\Omega^2-\gamma_\uparrow^2}$ with $2\Omega>\gamma_\downarrow$.

So, even though the experiment is not directly sensitive to the spin state, it is able to reveal the spin dynamics and determine its characteristic frequency. Supplemented with a histogram for the periods with no electron on the dot, we may also estimate the tunneling parameters.

In our particular simulations, the past quantum state analysis provides decisive information about the occupation of the dot, and the occupation dynamics is well resolved, but we note that a more general procedure allows extraction of the physical parameters, even when the inferred dynamics does not assign the state with certainty.

An optimal estimation of the Rabi frequency is thus made by Bayes rule, which assumes a prior, e.g., uniform, probability distribution for $\Omega$, and evolves a separate density matrix $\rho_\Omega(t)$ for each of a set of candidate values. The QPC current outcomes occur with higher probabilities for some than for other conditional states  $\rho_\Omega$, and the likelihood $L(\Omega)$ for the different $\Omega$ are merely multiplied by these probabilities at each time step. During the parallel time evolution of the different candidate density matrices, the value of $\Omega$ with the highest likelihood represents the best estimate for the Rabi frequency.
An example of the Bayes rule evolution of $L(\Omega)$ with time is shown in Fig. \ref{fig:BW}.
We see that the method quickly identifies the plausible range of values for $\Omega$, and that the estimate becomes more peaked as data accumulates. A more detailed analysis shows that the estimation error scales as $1/\sqrt{T}$ with long probing times $T$\cite{PhysRevA.87.032115}.

\begin{figure}
 \centering
\includegraphics[scale=0.28]{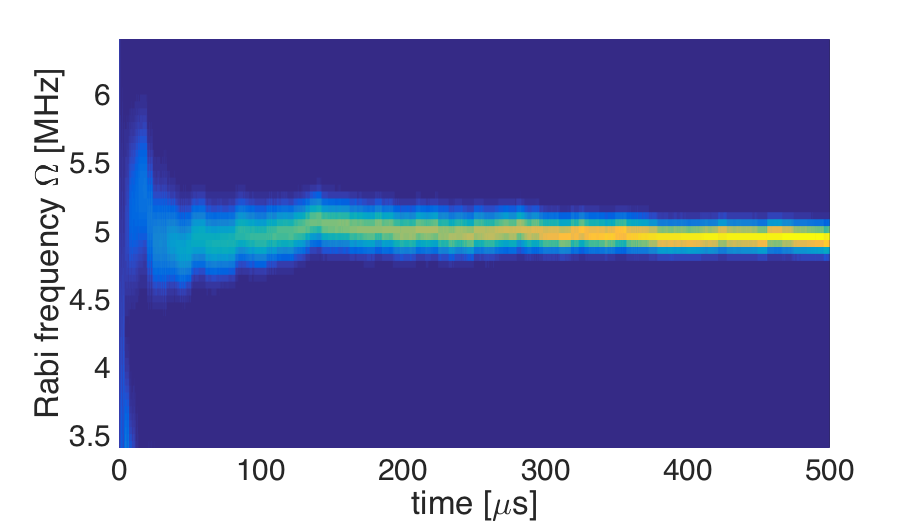}
 \caption{Bayesian estimation of Rabi frequency $\Omega$. The color scale depicts the likelihood function $L(\Omega)$. We used a grid of 60 candidate values between $3.5$ MHz and $6.5$ MHz. We see clear convergence towards the correct value of $\Omega=5$~MHz.} \label{fig:Bayes}
 \end{figure}

It is in principle possible to identify multiple parameters by Bayes method, but it requires propagation of conditional density matrices, with each parameter exploring a grid of candidate values. One therefore has recourse to other methods if the number of parameters is too high. For incoherent processes, the so-called parameter re-estimation method consists in determining the dynamics of the system subject to a choice of parameters, and then extracting the apparent rates from the dynamics. Using an initial guess for the rate parameters, one determines the time dependent occupation of the different states as well as the joint distribution for occupying state $i$ at time $t$, and state $j$ at the next time step according to the full measurement record. From the same probabilities and joint probabilities, one subsequently  extracts new candidate values for the transition probabilities. The whole data record is then analyzed again, but with these inferred rates, and the procedure is repeated until it has converged and the rates applied in the dynamics are in agreement with the rates inferred from the population dynamics.

This so-called Baum-Welch estimation scheme has been developed for classical parameter estimation, which means that it also applies to incoherent quantum processes as shown in case of atomic dynamics in a cavity QED experiment \cite{PhysRevA.89.043839}. It cannot be applied directly here, since we have a combination of coherent (Rabi frequency $\Omega$) and incoherent (tunneling rates $\gamma_{\uparrow,\downarrow}$) dynamics. In particular, the correlations between the population of the states at different times are not only due to the incoherent rates but also due to the coherent processes coupling the populations of states $i$ and $j$ via the coherence $\rho_{ij}$, $i\neq j$. Results of the modified Baum-Welch method designed to include the coherent part of the dynamics are shown in Fig. \ref{fig:BW}. The details of the method are discussed in Appendix \ref{AppBW}.

\begin{figure}
 \centering
 \includegraphics[scale=0.29]{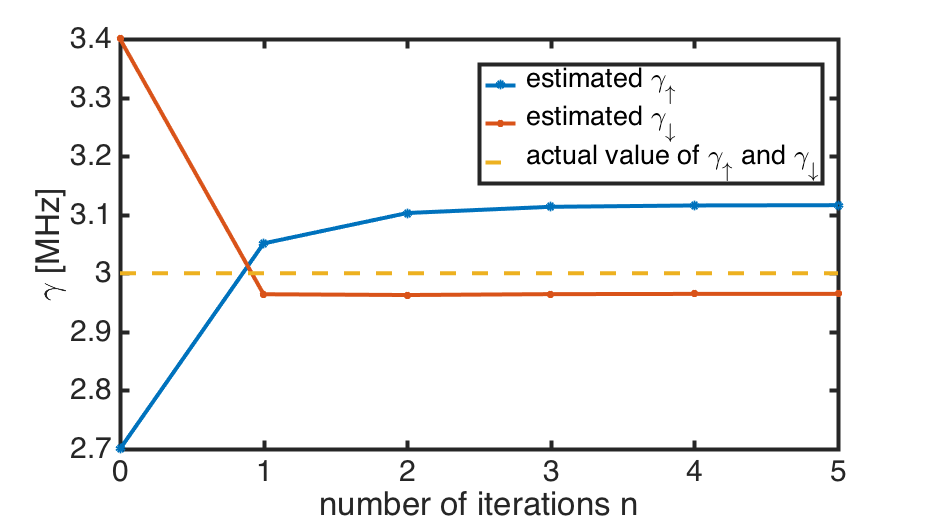}
 \caption{Estimation of $\gamma_\downarrow$, $\gamma_\uparrow$ using the modified Baum-Welch procedure. Assuming the correct Rabi frequency, $\Omega$, $\gamma_{\uparrow,\downarrow}$ is shown as a function of number of iterations $n$.}\label{fig:BW}
 \end{figure}

\begin{figure}
 \centering
  \includegraphics[scale=0.55]{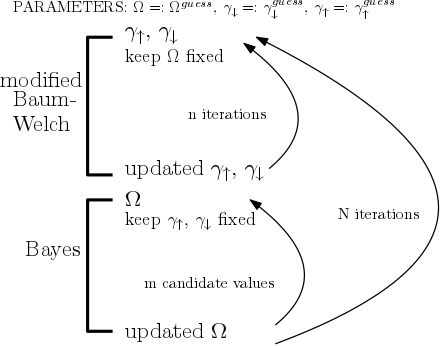}
 \caption{The scheme of our numerical approach to estimate both coherent and incoherent physical parameters: We fix the Rabi frequency and use $n$ iterations of the Baum-Welch re-estimation procedure to obtain consistent values of the rates. With these rates fixed, we use Bayesian estimation to pick the lost likely Rabi frequency among $m$ candidate values. The whole protocol is repeated $N$ times until all parameters have converged. }
 \label{fig:scheme}
 \end{figure}

Understanding how to estimate coherent parameters assuming known values of the incoherent ones and vice versa, we propose the  hybrid scheme illustrated in Fig. \ref{fig:scheme}.
The method assumes (guessed) values for the rates, and applies the Bayes rule method to find the most likely value of $\Omega$. That value, together with the rates is subsequently used in the propagation of the master equation for $\rho(t)$ and the adjoint equation for $E(t)$. With the values of $\rho(t)$ and $E(t+dt)$, we then calculate the joint probability for finding the system in state $i$ at $t$ and in $j$ at $t+dt$, which should agree on average with the rates assumed in the calculation of the dynamics. As illustrated in Fig. \ref{fig:scheme}, this effective Baum-Welch protocol can be iterated until convergence (n times) for the given $\Omega$, and the resulting rates are used in a new Bayes estimation of $\Omega$. The step is repeated (N times) until all parameters are converged and consistent with the data.

When applying this method to our simulated data $n=5$ modified Baum-Welch iterations and $N=5$ iterations of the full protocol were sufficient to estimate all three parameters. We observe in Fig.~\ref{fig:BW} that our estimation does not identify the `true' parameters used for the simulation exactly. This is due to the use of finite data sampling and better agreement is expected for longer probing times.

\section{Conclusions and Outlook}

In this article we have investigated the prospects of monitoring the spin and charge dynamics of a single electron quantum dot with a QPC.
We have shown that the conditional master equation allows inference of the quantum state of the system, and that analysis of the full temporal signal improves the statistical certainty about the tunneling events in the system.

In our model, the tunneling dynamics is correlated with the spin precession, and while the spin dynamics is not directly observable in the experiment, it is possible to infer its properties from the charge measurements. In particular, the Rabi frequency is evident in the distribution of time intervals where the quantum dot is charged. A Bayesian analysis of the noisy measurement current record permits a reliable estimate of $\Omega$, while a generalization of the classical Baum-Welch algorithm allows estimation of the tunneling rates.

By adjusting the bias across the charge sensor and therefore the strength of the measurement, our method offers a versatile approach to weak measurements of spin qubits~\cite{Jordan2007}. All of the ingredients of this approach have now been demonstrated experimentally~\cite{Hanson2007}. The key technology of spin readout via a charge sensor has been achieved both using dc current as in Fig. 1~\cite{Elzerman2004,Petta2180,QPCranieri,QPCcahay} as well as via the faster but conceptually similar technique of radio-frequency reflectrometry~\cite{Reilly2007,Cassidy2007,Ares2016,PhysRevX.7.011030,PhysRevLett.107.256804,PhysRevLett.108.046807,PhysRevB.88.125312}, to which our analysis can also be applied. Single-spin rotation can be achieved in several ways; by applying a local magnetic field~\cite{Koppens2006, Pla2012}, by electrically driven spin resonance~\cite{Nowack2007,Laird2007}, or by coupling to an on-chip field gradient~\cite{Pioro-Ladriere2008}.

Our simulations used realistic physical parameters. Thus, by operating at a qubit frequency of 40~GHz (corresponding to a magnetic field $\sim 7$~T in GaAs or $\sim 1.4$~T in Si), thermal fluctuations at $30$~mK may be neglected. The variation of our simulated QPC signal is compatible with the reported ability to perform charge readout and distinguish the occupation in a quantum dot in $100$~ns \cite{Barthel2010}. Finally, Rabi oscillations as slow as $1.5$~MHz \cite{Nowack2007} have been measured and may thus form comparison with the achievement of the method.

In spin quantum computing, one method for mitigating the fluctuating hyperfine coupling that limits operation fidelity is to track the instantaneous hyperfine field dynamically from its effect on the electron spin precession frequency~\cite{Klauser2006,Sergeevich2011,Shulman2014}. Also, a record of the spin qubit precession frequency can be used to estimate a local electric field~\cite{Dial2013}. 
We believe that our results can be applied in a number of such situations where parameters that govern spin evolution must be estimated efficiently.

\section*{Acknowledgments}
The authors acknowledge financial support from the Villum Foundantion Centre of Excellence, QUSCOPE, Royal Academy of Engineering, and EPSRC (EP/J015067/1).We thank Natalia Ares for stimulating discussions.

\appendix
\section{Quantum modified Baum-Welch method}
\label{AppBW}
The dynamics of a quantum system can be inferred from the measurement data via a stochastic master equation. The average inferred behavior may be at variance with the values assumed for the transition rates in the master equation. This forms the basis for the classical Baum-Welch parameter re-estimation algorithm which iteratively extracts the rates from the inferred dynamics and re-applies them in the inference process, until the values have converged.

The extraction of the rates is based on the estimation of the joint probability,
\begin{equation}
\label{eq:gamma}
C_t(i,j)=P(X_{t+dt}=j,X_t=i|s_1,...,s_N),
\end{equation}
that the system occupies the state $i$ at the time $t$ and the state $j$ at the next discrete time step $t+dt$ (conditioned on the overall measurement record $s_1,\dots, s_n$).
In the quantum case this joint classical probability has to be replaced by the probability that projective measurements at the two subsequent times yield the outcomes  $|i\rangle$ and $|j\rangle$, respectively. The quantum theory of measurements allows evaluation of this probability by taking into account the effect of the evolution of the system between such projective measurements. The past quantum state formalism \eqref{eq:genBorn}, readily generalizes to yield the expression
\begin{equation}
\label{eq:CorrKlaus}
C_t(i,j) = \tr{|j\rangle\langle j| e^{(L dt)}\{|i\rangle\langle i|\rho(t)|i\rangle\langle i|\}|j\rangle\langle j| E(t+dt)},
\end{equation}
where $\exp(L dt)$ is a superoperator denoting the propagator of the master equation for the short time step $dt$ between the (hypothetical) measurements on the system \footnote{This procedure estimates the entire transfer probability between the states, i.e., both contributions associated with the rate processes we want to estimate and contributions from unitary,  coherent processes. Due to the projection on the eigenstates in the expression \eqref{eq:CorrKlaus}, however,  coherent processes represented by  $\exp(L dt)$ will only contribute to second order in $dt$. We have verified this by omitting the coherent part of $\exp(L dt)$ in our calculations without any appreciable change of the result. }.
Now, the Baum-Welch procedure obtains the transition rate $\gamma_{ij}$  from the average of this correlation function over the entire measurement record, normalized to the average population of the initial state $|i\rangle$ \cite{press2007numerical, PhysRevA.89.043839}
\begin{equation}
\label{eq:condprob}
\gamma_{ij} dt=\frac{\sum_tC_t(i,j)}{\sum_t(\sum_jC_t(i,j))}.
\end{equation}

%

\end{document}